\documentclass[12pt,a4paper]{article}
\RequirePackage{amsmath,mathrsfs,amsfonts,dsfont}
\RequirePackage{braket}
\RequirePackage{tabularx}
\RequirePackage{nicefrac}
\usepackage{geometry,graphicx}
\RequirePackage{colortbl,booktabs,xcolor}
\RequirePackage[symbol]{footmisc}
\usepackage{makecell}
\usepackage{comment}
\usepackage{lscape}
\usepackage{blkarray}
\usepackage{amssymb}
\DeclareMathOperator{\Tr}{Tr}
\definecolor{redmar}{rgb}{.9,0,0}

\def\AEF{A.E. Faraggi}

\def\EJP#1#2#3{{\it Eur.\ Phys.\ Jour.}\/ {\bf C#1} (#2) #3}

\def\JHEP#1#2#3{{\it JHEP}\/ {\bf #1} (#2) #3}
\def\NPB#1#2#3{{\it Nucl.\ Phys.}\/ {\bf B#1} (#2) #3}
\def\PLB#1#2#3{{\it Phys.\ Lett.}\/ {\bf B#1} (#2) #3}
\def\PRD#1#2#3{{\it Phys.\ Rev.}\/ {\bf D#1} (#2) #3}
\def\PRL#1#2#3{{\it Phys.\ Rev.\ Lett.}\/ {\bf #1} (#2) #3}

\def\etal{{\it et al\/}}

\begin{document}
\begin{titlepage}
\samepage{
\setcounter{page}{1}
\rightline{}
\rightline{April 2019}

\vfill
\begin{center}
{\Large \bf{Doublet-Triplet Splitting in Fertile \\ \medskip 
Left-Right Symmetric Heterotic String Vacua}}

\vspace{1cm}
\vfill

{\large Alon E. Faraggi$^{1}$\footnote{E-mail address: alon.faraggi@liv.ac.uk},
Glyn Harries$^{1}$\footnote{E-mail address: g.harries@liv.ac.uk},\\
\medskip 
Benjamin Percival$^{1}$\footnote{E-mail address: benjamin.percival@liv.ac.uk} and John
Rizos$^{2}$\footnote{E-mail address: irizos@uoi.gr}}\\

\vspace{1cm}

{\it $^{1}$ Dept.\ of Mathematical Sciences,\\ 
University of Liverpool, Liverpool L69 7ZL, UK\\}

\vspace{.08in}

{\it $^{2}$ Department of Physics, University of Ioannina, GR45110 Ioannina,
Greece\\}

\vspace{.025in}
\end{center}

\vfill
\begin{abstract}
\noindent

Classification of Left--Right Symmetric (LRS) heterotic--string vacua
in the free fermionic formulation, using random generation of Generalised
GSO (GGSO) projection coefficients, produced phenomenologically viable 
models with probability $4\times 10^{-11}$. 
Extracting substantial
number of phenomenologically viable models requires 
modification of the classification method. 
This is achieved 
by identifying phenomenologically amenable conditions on the 
Generalised GSO projection coefficients that are randomly generated 
at the $SO(10)$ level. 
Around each of these fertile cores we perform a complete 
LRS classification, generating viable models with probabilility 
$1.4\times 10^{-2}$, hence increasing the probability of generating 
phenomenologically viable models by nine orders of magnitude, 
and producing some $1.4\times 10^5$ such models. 
In the process we identify a doublet--triplet selection mechanism 
that operates in twisted sectors of the string models that break
the $SO(10)$ symmetry to the Pati--Salam subgroup. This mechanism
therefore operates as well in free fermionic models with Pati--Salam and 
Standard--like Model $SO(10)$ subgroups.
\end{abstract}

\smallskip}
\end{titlepage}

\section{Introduction}
\normalsize
The Standard Model of particle physics agrees with all observational 
data to date. The discovery of a scalar resonance, compatible with the 
Standard Model Higgs particle, lends further support to the possibility that
the Standard Model provides viable parameterisation of all
experimental observables
up to the GUT or Planck scales. Further elucidation of the Standard Model
parameters can therefore only be obtained by fusing it with gravity, {\it i.e.}
in a theory of quantum gravity. String theory is the leading contemporary 
framework that enables the pursuit of this synthesis, as its consistency 
conditions mandate the existence of the gauge and matter structures that form
the bedrock of the Standard Model. This necessitates the construction 
of string models that are compatible with the Standard Model data
\cite{spreview}. 

An appealing feature of the Standard Model is the embedding of its 
matter states in chiral $SO(10)$ spinorial {\bf16} representations. 
This characteristic is reproduced in the heterotic $E_8\times E_8$ string
theory \cite{hetecan} that gives rise to chiral {\bf 16} $SO(10)$ 
representations in the perturbative spectrum. The construction
of phenomenological string models proceeds by studying compactifications
of the heterotic--string to four dimensions. Among the string 
models that reproduce a large number of phenomenological 
three generation models 
with $SO(10)$ embedding of the chiral spectrum are the heterotic--string 
models in the free fermionic formulation that correspond to 
$Z_2\times Z_2$ orbifold compactifications at special points in the 
moduli space with discrete Wilson lines \cite{z2xz2}.

Early constructions of phenomenological free fermionic models provided
isolated examples of three generation models with 
$SU(5)\times U(1)$ (FSU5) \cite{fsu5}, 
$SU(3)\times SU(2)\times U(1)^2$ (SLM) \cite{slm}, 
$SO(6)\times SO(4)$ (PS) \cite{alr},
and
$SU(3)\times U(1)\times SU(2)^2$ (LRS) \cite{lrs}
unbroken $SO(10)$ subgroups, and the canonical GUT 
embedding of the Standard Model weak hypercharge. 
Systematic computerised methods to classify large spaces of 
fermionic $Z_2\times Z_2$ orbifold models were developed 
over the past two decades, initially for the
type II superstring \cite{gkr}, 
and extended to the heterotic--string 
\cite{fknr, fkr, acfkr,frs, hasan, SLM, LRSGlyn}. 
The classification of vacua with unbroken $SO(10)$ subgroup 
revealed the existence of a new symmetry in the space
of heterotic--string compactifications with $(2,0)$ worldsheet supersymmetry, 
dubbed spinor--vector duality \cite{fkr, cfkr, panos}. The classification 
method provides an efficient algorithm to extract string vacua with specific 
phenomenological properties, leading to the discovery of: exophobic string 
vacua \cite{acfkr}; heterotic--string models with unbroken 
$SU(6)\times SU(2)$ gauge group \cite{su62}; and 
the construction of string vacua 
with an extra $Z^\prime$ compatible with all the low scale constraints
\cite{frzprime}. 
We note that computerised analysis of large sets of string vacua have been 
performed by other research groups \cite{statistical}.

The systematic heterotic--string classification is a progressive program. 
It was initially performed
for vacua with unbroken $SO(10)$ with respect to the spinorial
and anti--spinorial representations \cite{fknr}. 
It was extended to include vectorial representations \cite{fkr}, 
and subsequently to include all matter representations arising in the 
string models with 
$SO(6)\times SO(4)$  (PS) \cite{acfkr}; 
$SU(5)\times U(1)$  (FSU5) \cite{frs, hasan}; 
$SU(3)\times SU(2)\times U(1)^2$ (SLM) \cite{SLM}; 
and 
$SU(3)\times U(1)\times SU(2)^2$ (LRS) \cite{LRSGlyn}, 
unbroken $SO(10)$ subgroups, with each step representing an
increase in complexity. The case of PS models utilise 
solely RNS boundary conditions \cite{acfkr}, whereas the three other cases 
utilise RNS and complex boundary conditions.
The FSU5 models utilise a single basis vector that breaks the 
$SO(10)$ gauge symmetry \cite{frs, hasan}, 
whereas the SLM models necessarily include 
two such basis vectors \cite{SLM}.
The SLM models therefore contain a proliferation of sectors
that include an $SO(10)$ breaking vector and produce exotic states. 
The result is that the frequency of viable three generation 
models in the total space of models is reduced, making
the random based classification method inefficient. To circumvent
this problem fertile conditions have been identified that facilitate
the extraction of three generations SLM vacua with varying 
phenomenological characteristics \cite{SLM}.
We remark that the genetic algorithm developed in ref. \cite{ar}
provides an alternative method to extract vacua with phenomenological 
characteristics, albeit not to classify large classes of them. 
We further note that employing fertility condition analysis 
of string vacua is also adopted in analysis of other classes of 
string vacua \cite{vaudrevange}. 

The situation in the case of the LRS models is similar to that 
of the SLM models, with the added complexity that the LRS models 
do not admit the $E_6$ embedding of the charges in the 
extension of $SO(10)\times U(1)$ to $E_6$ \cite{lrs}. 
While the LRS models can be constructed with a single $SO(10)$ basis vector
$\alpha$, the vector $2\alpha$ breaks the $SO(10)$ symmetry as well 
\cite{LRSGlyn}. Thus, exotic states producing 
sectors arise in the LRS string models from basis vector combinations
with the vectors $\alpha$ and $2\alpha$ resulting again in 
proliferation of exotic producing sectors and diminishing the 
frequency of viable three generation vacua. A remedy to this
situation is provided by identifying a set of fertile conditions 
in the space of LRS free fermionic heterotic--string vacua. 

In this paper we undertake this task. In the process we 
uncover a doublet--triplet mechanism in the twisted 
sectors of the heterotic--string vacua. At the $SO(10)$ 
level vectorial {\bf 10} representations arise from the 
untwisted and twisted sectors. These decompose as
$5+{\bar 5}$ under $SU(5)$ and as $(3,1,1)+({\bar 3},1,1) + (1,2,2)$
under the LRS subgroup $SU(3)\times U(1)\times SU(2)^2$. In the 
case of the untwisted states a doublet--triplet splitting 
mechanism has been identified in PS, SLM and LRS string vacua that utilise 
asymmetric boundary conditions \cite{doublettriplet}. 
However, the free fermionic 
systematic classification method utilises symmetric 
boundary conditions. In PS, SLM and LRS string vacua with 
symmetric boundary conditions the untwisted sector produces
three pairs of colour triplets rather than electroweak 
doublets. In this paper we identify a doublet--triplet 
splitting mechanism in terms of the discrete torsions that 
appear in the one--loop partition function of the models and
that operates in the twisted sectors of the LRS models. The 
core of our fertility conditions revolve about the doublet--triplet
splitting mechanism in the twisted sectors, thus increasing 
the frequency of models that contain heavy and light Higgs 
representations.

As in the case of the SLM models, the classification is performed in two 
stages. The fertility conditions include GGSO phases that involve only 
basis vectors that do not break the $SO(10)$ GUT symmetry. Thus, the fertility 
conditions are implemented by a random search for $SO(10)$ vacua 
that satisfy these conditions, resulting in 19374 fertile cores. 
To these cores we add the $SO(10)$ breaking basis vector and 
generate a complete classification of LRS string vacua, generating 
some $9.92\times 10^6$ models from which $1.4\times 10^5$ 
satisfy all our phenomenological criteria. 
This result exceeds the random classification method 
of \cite{LRSGlyn} by four orders of magnitude in about 1/10 
computational time on a computer platform of similar power. 

Our paper is organised as follows: 
in section \ref{lrsffm} we discuss the general structure of the free
fermionic LRS models; section \ref{Fertility} summarises the fertility 
conditions employed in the analysis; in section \ref{CRAL} we discuss
the results of the analysis and in section \ref{dtsd} we introduce the 
doublet--triplet splitting mechanism that operates in the twisted 
sectors of the LRS models; in section \ref{exmodel} we analyse an 
exemplary model in some more detail; section \ref{conclusion} 
concludes our paper. 

\section{Left Right Symmetric Free Fermionic Models}\label{lrsffm}

This paper utilises the free fermionic formulation \cite{fff} of the heterotic 
string to explore the space of string vacua which possess the
Left-Right Symmetric (LRS) subgroup of $SO(10)$. The classification of such 
vacua was performed in \cite{LRSGlyn}. The models are constructed by defining 
a set of basis vectors and the Generalised Gliozzi-Scherk-Olive (GGSO) 
projection coefficients of the one-loop partition function. An overview 
is outlined in the following section but more details of the LRS 
classification can be found in \cite{LRSGlyn}.

In order to obtain LRS vacua, the $SO(10)$ GUT symmetry is broken directly 
at the string scale and the unbroken LRS subgroup of $SO(10)$ in the low 
energy effective field theory is 
$SU(3)_C\times U(1)_C \times SU(2)_L \times SU(2)_R$. 
Resulting models obey $N=1$ spacetime
supersymmetry and preserve the $SO(10)$ embedding of the weak hypercharge. 
Fermionic matter representations of the Standard Model are found in the 
spinorial \textbf{16} representation of $SO(10)$ decomposed under the
unbroken $SO(10)$ subgroup. Similarly, vectorial representations, 
including the Standard Model Light Higgs, derive from  the \textbf{10} 
representation of $SO(10)$.

\subsection{The Free Fermionic Formulation}
In this section, a brief overview of the the free fermionic formulation 
will be outlined. We will also draw attention to key features relevant for 
the discussion of fertile regions and doublet-triplet splitting in the LRS 
models. 

In the free fermionic formulation, the heterotic string is formulated directly 
in four space-time dimensions and the extra degrees of freedom needed to 
cancel the conformal anomaly are interpreted as free fermions propagating on 
the two dimensional
string worldsheet. Within the lightcone gauge, this results in having 20 
left--moving and
44 right--moving free fermions. In the standard notation, the left movers are 
represented by
$\psi^{\mu}_{1,2} \; , \; \chi^{1,\ldots,6} \; , \; y^{1,\ldots,6}\; , \;
w^{1,\ldots,6}$ and the right movers by $\overline{y}^{1,\ldots,6}\; ,
\;\overline{w}^{1,\ldots,6}\; , \;\overline{\psi}^{1,\ldots,5}\; ,
\;\overline{\eta}^{1,2,3}\; , \; \overline{\phi}^{1,\ldots,8}$. 
The six compactified directions of the internal manifold correspond to the 
$\{y,w|\bar{y},\bar{w}\}^{1,...,6}$, while the $\bar{\psi}^{1,...,5}$ generate 
the $SO(10)$ GUT and the $\bar{\phi}^{1,...,8}$ generate the hidden sector 
$SO(16)$ group. 

A free fermionic string model is defined through boundary condition basis 
vectors that specify the transformational properties of the free fermions as 
they propagate around the two non-contractible loops of the one--loop 
partition function. These basis vectors are 64-dimensional and are of the 
form:
\begin{equation*}
v_i = \{ \alpha_i(f_1), \ldots, \alpha_i(f_{20}) \; | \;
\alpha_i(\overline{f}_1) , \ldots, \alpha_i(\overline{f}_{44}) \}, \ \ \ 
i=1,...,64
\end{equation*}
where the boundary condition of a fermion, 
$\alpha(f)$, is defined through:
\begin{equation*}
f_j \rightarrow -e^{i\pi \alpha_i (f_j)} f_j \qquad j = 1,\ldots,64
\end{equation*}
so that $\alpha(f)=0,1$ correspond to real boundary conditions and 
$\alpha(f)=\frac{1}{2}$ corresponds to a complex boundary condition.

A model is constructed with two ingredients. First, is a set of basis vectors 
$v_{i=1,...,k}$, which span a space $\Xi$ of all linear combinations, 
$\alpha$, which we call sectors. Second, is a set of distinct GGSO 
projection coefficients $C\binom{v_i}{v_j}$, where $i>j$ due to modular 
invariance consistency conditions leaving $2^{N(N-1)/2}$ independent 
coefficients. 

With these two ingredients, we can construct the modular invariant 
Hilbert space $\mathcal{H}$ of states $\ket{S_\alpha}$ of the model 
through the one-loop GGSO projection such that:
\begin{equation}
    \mathcal{H}=\bigoplus_{\alpha\in\Xi}\prod^{k}_{i=1}
\left\{ e^{i\pi v_i\cdot F_{\alpha}}\ket{S_\alpha}=\delta_{\alpha}
C\binom{\alpha}{v_i}^*\ket{S_\alpha}\right\}\mathcal{H}_\alpha
\end{equation}
where $F_\alpha$ is the fermion number operator and $\delta_\alpha=1,-1$ 
is the spin-statistics index.
\subsection{Left-Right Symmetric Models}
Before specialising to the LRS case, we first construct $SO(10)$ models. 
We use a set of 12 basis vectors that are common to those used in recent 
free fermionic classifications
\cite{acfkr, frs, SLM, 421}:
\begin{eqnarray}\label{basis12}
v_1={\mathds{1}}&=&\{\psi^\mu,\
\chi^{1,\dots,6},y^{1,\dots,6}, \omega^{1,\dots,6}| \nonumber\\
& & ~~~\overline{y}^{1,\dots,6},\overline{\omega}^{1,\dots,6},
\overline{\eta}^{1,2,3},
\overline{\psi}^{1,\dots,5},\overline{\phi}^{1,\dots,8}\},\nonumber\\
v_2=S&=&\{{\psi^\mu},\chi^{1,\dots,6}\},\nonumber\\
v_{2+i}={e_i}&=&\{y^{i},\omega^{i}\; | \; \overline{y}^i,\overline{\omega}^i\},
\
i=1,\dots,6,\nonumber\\
v_{9}={b_1}&=&\{\chi^{34},\chi^{56},y^{34},y^{56}\; | \; \overline{y}^{34},
\overline{y}^{56},\overline{\eta}^1,\overline{\psi}^{1,\dots,5}\},\\
v_{10}={b_2}&=&\{\chi^{12},\chi^{56},y^{12},y^{56}\; | \; \overline{y}^{12},
\overline{y}^{56},\overline{\eta}^2,\overline{\psi}^{1,\dots,5}\},\nonumber\\
v_{11}=z_1&=&\{\overline{\phi}^{1,\dots,4}\},\nonumber\\
v_{12}=z_2&=&\{\overline{\phi}^{5,\dots,8}\}.
\nonumber
\end{eqnarray}
where the fermions which appear in the basis vectors have periodic 
(Ramond) boundary conditions, whereas those not included have antiperiodic 
(Neveu-Schwarz) boundary conditions.

The untwisted vector bosons present due to this choice of basis vectors
generate the gauge group $SO(10) \times U(1)^3 \times SO(8)^2$ in the 
adjoint representation.

A key role is played by the vectors $b_1$ and $b_2$ in these models as they 
define the $SO(10)$ gauge symmetry and correspond to $Z_2\times 
Z_2$ orbifold twists which break the $N=4$ supersymmetry, 
obeyed by the other 10 vectors, to $N=1$. The third twisted sector 
is given by the linear combination $b_3=b_1+b_2+x$, where the $x$ vector is 
the combination:
\begin{equation}\label{x}
    x=1+S+\sum^6_{i=1}e_i+\sum^2_{k=1} z_k=\{\bar{\eta}^{123},\bar{\psi}^{12345}\}.
\end{equation}
This vector plays an important role in these models as a map from spinorial 
\textbf{16} sectors of $SO(10)$ to vectorial \textbf{10} sectors. 

In order to break the $SO(10)$ models down to the LRS subgroup we add a single 
breaking basis vector:
\begin{equation}
v_{13}=\alpha = \{ \overline{\psi}^{1,2,3} = \frac{1}{2} \; , \;
\overline{\eta}^{1,2,3} = \frac{1}{2}\; , \; \overline{\phi}^{1,\ldots,6} =
\frac{1}{2}\; , \; \overline{\phi}^7 \}
\end{equation}
which will leave the unbroken $SO(10)$ subgroup $SU(3)\times SU(2)^2\times U(1)$. 
\subsection{GGSO Projections}
The next ingredient of the 
free fermionic models are the GGSO projection coefficients 
$C\binom{v_i}{v_j}$. 

Since we have 13 basis vectors, our GGSO coefficients span a $13\times 13$ 
matrix. Due to modular invariance constraints, the lower triangle of the 
matrix containing 78 coefficients are fixed by the upper triangle. 
Modular Invariance constraints also lead to the following demands on the 
leading diagonal phases:
\begin{equation}\label{GSOdiagonals}
\begin{split}
 C\binom{e_i}{e_i} &= -C\binom{e_i}{\mathds{1}}\qquad i=1,\ldots, 6\\
 C\binom{b_k}{b_k} &= C\binom{b_k}{\mathds{1}}\qquad k=1,2\\
 C\binom{z_k}{z_k} &= C\binom{z_k}{\mathds{1}}\qquad k=1,2\\
 C\binom{\alpha}{\alpha} &= C\binom{\alpha}{\mathds{1}}
\end{split}
\end{equation}
The matrix entries are further constrained through imposing 
$N = 1$ supersymmetry. This can be done by requiring:
\begin{equation}
C\binom{\mathds{1}}{\mathds{1}}=C\binom{S}{\mathds{1}} = C\binom{S}{S} =
C\binom{S}{e_i} = C\binom{S}{b_k} = C\binom{S}{z_k} = C\binom{S}{\alpha} = -1
\end{equation}
where $i=1,\ldots,6$ and $k=1,2$. All these constraints leave us with 66 
independent coefficients and therefore $2^{66} \approx 7.38\times 10^{19}$ 
distinct LRS string vacua. \\ \\ 
This is too large a space to explore with a computer program and so in 
\cite{LRSGlyn} a sample of $10^{11}$ was explored and found to produce 
phenomenologically viable vacua with probability $4\times 10^{-11}$. 
This tiny probability is the key motivation for modifying this 
classification procedure 
through the use of imposing phenomenological constraints 
in the smaller space of $SO(10)$ 
models. This corresponds to constraining the GGSO coefficients relating 
to the first 12 basis vectors, which form a
$12\times 12$ matrix. 
\subsection{Properties of the String Spectrum}
The sectors in the model can be characterised according to the left and 
right moving vacuum separately. Physical states must however satisfy 
the Virasoro matching condition:
\begin{equation}
    M_L^2=-\frac{1}{2}+\frac{\xi_L \cdot\xi_L}{8}+N_L=-1
+\frac{\xi_R \cdot\xi_R}{8}+N_R=M_R^2
\end{equation}
where $N_L$ and $N_R$ are sums over left and right moving oscillators, 
respectively. In our models, sectors which have the products 
$\xi_L\cdot \xi_L = 0$ and $\xi_R\cdot\xi_R = 0,4,6,8$ can produce
spacetime vector bosons, which
determine the gauge symmetry in a given vacuum. We note that only 
massless states are phenomenologically interesting as 
massive states will be at scales comparable to the Plank mass. 

From the untwisted sector vector bosons we obtain a full gauge group of:
\begin{align}\label{GG}
\begin{split}
\text{Observable}: \ \ \ &SU(3)_C \times U(1)_C \times SU(2)_L \times SU(2)_R
\times U(1)_{1,2,3} \\ 
\text{Hidden}: \ \ \  &SU(4) \times U(1)_4 \times SU(2) \times U(1)_5 \times U(1)_7 \times U(1)_8
\end{split}
\end{align}
where the weak hypercharge is given by:
\begin{equation}
U(1)_Y = \frac{1}{3}U(1)_C + \frac{1}{2}U(1)_L,
\end{equation}
such that $U(1)_C =
\frac{3}{2}U(1)_{B-L}$ and $U(1)_L = 2U(1)_{T_{3_R}}$.

In order to obtain the charge $Q$ associated to a $U(1)$ 
current generated by a fermion $f$ we use:
\begin{equation}
    Q(f)=\frac{1}{2}\alpha(f)+F(f)
\end{equation}
where $\alpha(f)$ is the boundary condition of the fermion 
in the sector and $F(f)$ is the fermion number given by:
\begin{subequations}
\begin{equation}
F (f)=\begin{cases}+1&\text{for}\; f\\ -1&\text{for}\; f^* \end{cases}
\end{equation}
for fermionic oscillators and their complex conjugates, whereas for degenerate Ramond vacua it is:
\begin{equation}
\begin{split}
& F \ket{+}_R = 0 \\
& F \ket{-}_R = -1,
\end{split}
\end{equation}
\end{subequations}
where $\ket{+}_R = \ket{0}$ is a degenerated vacuum with no 
oscillator and $\ket{-}_R = f_0^{\dagger}\ket{0}$ is the
degenerated vacua with one zero mode oscillator.
\section{Fertility Conditions}\label{Fertility}
In order to narrow the search of the $2^{66}$ vacua on phenomenologically 
promising regions, we examine the GGSO coefficients at the $SO(10)$ level, 
which means a $12\times 12$ matrix with 55 independent coefficients. 
The aim of this section is to apply further constraints that we 
call `fertility conditions' on the $SO(10)$ models. The models 
satisfying the fertility conditions we call `fertile cores'.  
The conditions are chosen so as to increase the likelihood of finding 
phenomenologically viable vacua at the LRS level.

After obtaining fertile cores we perform a comprehensive classification 
of all models resulting from the cores by iterating over all $\alpha$ 
projection coefficients values. This methodology was used with great 
success in \cite{SLM} where phenomenologically viable standard-like 
vacua were found in great abundance through the use of fertility conditions. 

\subsection{Observable Spinorial Sectors}\label{osconstraints}
The choice of basis vectors in equation (\ref{basis12}) means that sectors giving rise to
states of a particular representation of the gauge group can be written 
compactly as a function of $p,q,r,s=0,1$. These 16 possibilities correspond 
to the 16 fixed points of each twisted plane of the 
$Z_2\times Z_2$ orbifold. For example, 
the observable $SO(10)$ spinorial sectors are:
\begin{eqnarray}
B_{pqrs}^{(1)} &=& S + b_1 + pe_3 + qe_4 + re_5 + se_6
\nonumber \\ 
&=& \{\psi^{\mu},\chi^{1,2},(1-p)y^3\bar{y}^3,
pw^3\bar{w}^3,(1-q)y^4\bar{y}^4,qw^4\bar{w}^4,
 \\
& & ~~~ (1-r)y^5\bar{y}^5,rw^5\bar{w}^5,(1-s)y^6\bar{y}^6,
sw^6\bar{w}^6,\bar{\eta}^{1},\bar{\psi}^{1,\ldots,5}\}
\nonumber \\
B_{pqrs}^{(2)} &=&  S + b_2 + pe_1 + qe_2 + re_5 + se_6\nonumber \\
B_{pqrs}^{(3)} &=&  S + b_3 + pe_1 + qe_2 + re_3 + se_4\nonumber
\end{eqnarray}
where $p,q,r,s = 0,1$ and $b_3 = b_1 + b_2 + x$. These 48 sectors contain the
$\textbf{16}$ and $\overline{\textbf{16}}$ spinorial representations of the
$SO(10)$.

With this information we can begin classifying the spinorial/antispinorials, 
$\textbf{16}$/$\overline{\textbf{16}}$, of $SO(10)$. 
The spinorials/antispinorial can be determined to give rise to either 
left or right chirality states, leaving 4 classification numbers: 
$N_L,\overline{N}_L,N_R,\overline{N}_R$.
To determine whether a sector gives rise to a spinorial or 
antispinorial, we inspect the projectors on $B^A, \ A=1,2,3$: 
\begin{eqnarray}\nonumber
P^1_{pqrs} &=& \frac{1}{2^4}\prod_{i=1,2}\left[1-C\begin{pmatrix}
B^1_{pqrs}\\
e_i
\end{pmatrix}^*\right]\prod_{a=1,2}\left[1-C\begin{pmatrix}
B^1_{pqrs}\\
z_a
\end{pmatrix}^*\right]\\
P^2_{pqrs} &=& \frac{1}{2^4}\prod_{i=3,4}\left[1-C\begin{pmatrix}
B^2_{pqrs}\\
e_i
\end{pmatrix}^*\right]\prod_{a=1,2}\left[1-C\begin{pmatrix}
B^2_{pqrs}\\
z_a
\end{pmatrix}^*\right]\\
\nonumber
P^3_{pqrs} &=& \frac{1}{2^4}\prod_{i=5,6}\left[1-C\begin{pmatrix}
B^3_{pqrs}\\
e_i
\end{pmatrix}^*\right]\prod_{a=1,2}\left[1-C\begin{pmatrix}
B^3_{pqrs}\\
z_a
\end{pmatrix}^*\right]
\end{eqnarray}
and the chirality phases:
\begin{eqnarray}
X^1_{pqrs} &=& -C\begin{pmatrix}\nonumber
B^1_{pqrs}\\
S+b_2+(1-r)e_5+(1-s)e_6
\end{pmatrix}^*\\
X^2_{pqrs} &=& -C\begin{pmatrix}
B^2_{pqrs}\\
S+b_1+(1-r)e_5+(1-s)e_6
\end{pmatrix}^*\\
X^3_{pqrs} &=& -C\begin{pmatrix}\nonumber
B^3_{pqrs}\\
S+b_1+(1-r)e_3+(1-s)e_4
\end{pmatrix}^*
\end{eqnarray}
which together let us define $N_{16},N_{\overline{16}}$ as:
\begin{eqnarray}
N_{16} &=& \frac{1}{2}\sum_{\substack{A=1,2,3 \\ p,q,r,s=0,1}} 
P_{pqrs}^A\left(1 + X^A_{pqrs}\right) \\
N_{\overline{16}} &=& \frac{1}{2}\sum_{\substack{A=1,2,3 \\ p,q,r,s=0,1}} 
P_{pqrs}^A\left(1 - X^A_{pqrs}\right) \nonumber
\end{eqnarray}
The number of spinorials/anti-spinorials alone is not sufficient to 
describe the phenomenological properties of the models under consideration as we need to consider what happens as the $SO(10)$ GUT is broken. 

Recall that the basis vector $v_{13} = \alpha$ induces $SO(10)$ gauge symmetry breaking.
The spinorial representations of $SO(10)$, are decomposed under the residual 
$SU(3)_C \times U(1)_C \times
SU(2)_L \times SU(2)_R$ gauge group as:
\begin{align}
\label{sp_deco}
\mathbf{16} &= Q_L \left(\textbf{3} , +\frac{1}{2}, \textbf{2} , \textbf{1}\right)  + Q_R\left(\overline{\textbf{3}} , -\frac{1}{2}, \textbf{1} , \textbf{2}\right) + L_L\left(\textbf{1} , -\frac{3}{2}, \textbf{2} , \textbf{1}\right) + L_R\left(\textbf{1} , +\frac{3}{2}, \textbf{1} , \textbf{2}\right)\\
\overline{\mathbf{16}} &= \overline{Q}_L \left(\overline{\textbf{3}} , -\frac{1}{2}, \textbf{2} , \textbf{1}\right)  + \overline{Q}_R\left({\textbf{3}} , +\frac{1}{2}, \textbf{1} , \textbf{2}\right) + \overline{L}_L\left(\textbf{1} , +\frac{3}{2}, \textbf{2} , \textbf{1}\right) + \overline{L}_R\left(\textbf{1} , -\frac{3}{2}, \textbf{1} , \textbf{2}\right)
\nonumber
\end{align}
Only one of the spinorial components $Q_L, Q_R, L_L, L_R$ survives the $\alpha$ projections. The same is true for the anti-spinorials. That is, in order to accommodate the fields of one fermion generation we need at least 
four $SO(10)$ spinorials and properly adjusted projections. This poses a challenge to any computer-based model scan. A lot of 
computer time is allocated in examining unacceptable incomplete generation models. 

Remarkably, there is a way of partially overcoming this important problem. It turns out that the GGSO projection of the vector $2\alpha + x$
when acting on spinorials differentiates between left and right states. 
In addition, as dictated by $C\binom{v_i}{v_j}$ properties, this projection does not act on the GGSO phases associated to
the $\alpha$ vector.
Indeed, the GGSO projection of the vector $2\alpha + x=
\{\bar{\psi}^{45},\bar{\phi}^{1,\dots,6}\}$ gives:
\begin{align}\label{2apxprojection}
\begin{split}
 e^{i\pi (2\alpha+x)\cdot F_{B^A_{pqrs}}}\ket{\left({B^A_{pqrs}}\right)_0} &=
\delta_{B^A_{pqrs}}C\left(B^A_{pqrs}\atop 2\alpha+x\right)^*\ket{\left({B^A_{pqrs}}\right)_0}\\
\Rightarrow\text{ch}\left(\bar{\psi}^{4,5}\right) &=
-C\left(B^A_{pqrs}\atop x\right)
\end{split}
\end{align}
where we have used that $2\alpha + x \cap B^{(A)}_{pqrs}= \{\bar{\psi}^{45}\}$ and the notation $\text{ch}\left(\bar{\psi}^{4,5}\right)$ stands for the $SO(4)\sim {SU(2)}_L\times{SU(2)}_R$ chirality. In other words, the $2\alpha + x$ projection selects between left and right states. 
Adopting the convention:
\begin{align}
C\left(B^A_{pqrs}\atop x\right) = 
\begin{cases} 
+1 \ \leftrightarrow \ \text{Left}\ (Q_L + L_L)\\
-1 \ \leftrightarrow \ \text{Right}\ (Q_R + L_R)\\
\end{cases}
\end{align}
we can write analytic formulas for the number
of left spinorials, $N_L$, right spinorials, $N_R$, as well as  
the left and right anti-spinorials $\overline{N}_L,\overline{N}_R$ respectively:
\begin{eqnarray}\label{SpinorialNumbers}
N_{L} &=& \frac{1}{4}\sum_{\substack{A=1,2,3 \\ p,q,r,s=0,1}} 
P_{pqrs}^A\left(1 + X^A_{pqrs}\right)\left[1 + C\begin{pmatrix}
B^A_{pqrs}\\
x
\end{pmatrix}\right] \\
N_{R} &=& \frac{1}{4}\sum_{\substack{A=1,2,3 \\ p,q,r,s=0,1}} 
P_{pqrs}^A\left(1 + X^A_{pqrs}\right)\left[1 - C\begin{pmatrix}
B^A_{pqrs}\\
x
\end{pmatrix}\right] \\
\overline{N}_{L} &=& \frac{1}{4}\sum_{\substack{A=1,2,3 \\ p,q,r,s=0,1}} 
P_{pqrs}^A\left(1 - X^A_{pqrs}\right)\left[1 + C\begin{pmatrix}
B^A_{pqrs}\\
x
\end{pmatrix}\right]\\
\overline{N}_{R} &=& \frac{1}{4}\sum_{\substack{A=1,2,3 \\ p,q,r,s=0,1}} 
P_{pqrs}^A\left(1 - X^A_{pqrs}\right)\left[1 - C\begin{pmatrix}
B^A_{pqrs}\\
x
\end{pmatrix}\right]
\end{eqnarray}
An eventually complete generation $SO(10)$ configuration should satisfy
\begin{align}
N_L - \overline{N}_L = {N}_R - \overline{N}_R \ge 2\,n_g
\label{fc_spin}
\end{align}
where $n_g$ stands for the number of generations. The factor of two in the last equation is necessary in order to compensate for the additional truncation imposed by the $\alpha$ vector projection. 

Furthermore, a consistent low energy model should include Higgs fields to break the $SU(3)_C\times{U(1)}_C\times{SU(2)}_L\times{SU(2)}_R$ gauge symmetry to that of the Standard Model. 
The necessary states, referred to as heavy Higgs fields 
transform as right-handed doublets
\begin{align}
 L_R^H\left(\textbf{1} , +\frac{3}{2}, \textbf{1} , \textbf{2}\right) + \overline{L}_R^H\left(\textbf{1} , -\frac{3}{2}, \textbf{1} , \textbf{2}\right) 
\end{align}
and lie in an additional pair of spinorial/anti-spinorial ($\mathbf{16}/ \overline{\mathbf{16}}$) $SO(10)$ representations. This leads to the additional constraint
\begin{align}
{N}_R > 2\,n_g. \label{fc_spinb}
\end{align}
We will refer to \eqref{fc_spin},\eqref{fc_spinb} as fertility conditions regarding spinorials. $SO(10)$ configurations enjoying this property 
are most likely to end up in phenomenologically viable Left-Right Symmetric Models when the $\alpha$ vector projection is also applied.

\subsection{Observable Vectorial Sectors and 
Doublet--Triplet Splitting }\label{ovsanddts}
Vectorials of $SO(10)$ gauge symmetry are of great importance to phenomenology for they accommodate the light Standard Model Higgs doublets.
In the class of models under consideration, massless $SO(10)$ vectorial states arise from the sectors 
\begin{eqnarray}
V_{pqrs}^{(I)} &=&  B_{pqrs}^{(I)}+x\ ,\ I=1,2,3.
\label{ovsinovs}
\end{eqnarray}
which contain four periodic right-moving complex fermions and consequently they admit one Neveu-Schwarz right-moving fermionic
oscillator $\left(\overline{\psi}^a_{1/2}/\overline{\psi}^{*a}_{1/2}\,,\,a = 1,\dots,5\right)$.
The vectorial representation of $SO(10)$ is decomposed under 
$SU(3)_C \times U(1)_C \times
SU(2)_L \times SU(2)_R$ as:
\begin{align}
\textbf{10} = d'\left(\textbf{3} , -\frac{1}{3}, \textbf{1} , \textbf{1}\right) + 
{d^{c}}'\left(\overline{\textbf{3}} , +\frac{1}{3},
\textbf{1} , \textbf{1}\right) + h\left(\textbf{1} , \textstyle{0}, \textbf{2} , 
\textbf{2}
\vphantom{+{\frac{1}{3}}^2} 
\right)
\label{vec_deco}
\end{align}
where the colored triplets are generated by the 
$\overline{\psi}^{1,2,3}_{1/2}/\overline{\psi}^{* 1,2,3}_{1/2}$ and 
the bi-doublet is generated by 
$\overline{\psi}^{4,5}_{1/2}/\overline{\psi}^{*4,5}_{1/2}$ oscillators.

At the $SO(10)$ level, i.e. taking into account 
GGSO projectors associated to $v_1,\dots, v_{12}$ vectors,
the total number of surviving vectorials, $N_{10}$, is given by
\begin{eqnarray}
N_{10}=\sum_{\substack{A=1,2,3 \\ p,q,r,s=0,1 }}R^A_{pqrs}
\end{eqnarray}
where
\begin{eqnarray}\label{VectProjector}
\nonumber R^{(1)}_{pqrs} &=& \frac{1}{2^4}\prod_{i=1,2}\left[1 - C\begin{pmatrix}
e_i \\
V^{(1)}_{pqrs}
\end{pmatrix}\right]
\prod_{a=1,2}\left[1- C\begin{pmatrix}
z_a \\
V^{(1)}_{pqrs}
\end{pmatrix}\right]\\
R^{(2)}_{pqrs} &=& \frac{1}{2^4}\prod_{i=3,4}\left[1-C\begin{pmatrix}
e_i \\
V^{(2)}_{pqrs}
\end{pmatrix}\right] 
\prod_{a=1,2}\left[1 - C\begin{pmatrix}
z_a \\
V^{(2)}_{pqrs}
\end{pmatrix}\right]\\
\nonumber R^{(3)}_{pqrs} &=& \frac{1}{2^4}\prod_{i=5,6}\left[1-C\begin{pmatrix}
e_i \\
V^{(3)}_{pqrs}
\end{pmatrix}\right] 
\prod_{a=1,2}\left[1 - C\begin{pmatrix}
z_a \\
V^{(3)}_{pqrs}
\end{pmatrix}\right].
\end{eqnarray}
However, the full GGSO projections, include also the gauge symmetry 
breaking $\alpha$ vector projections associated to 
$C\binom{\alpha}{v_i}\,,i=1,\dots,12$ phases. 
These projections act differently on the three states 
in \eqref{vec_deco}. As a result, only one of the vectorial 
segments (triplet, anti-triplet or bi-doublet) survives. 
Depending on the phase configuration, the $\alpha$ 
related projections can eliminate all Standard 
Model doublets leading to unacceptable phenomenology. 
The mere existence of $SO(10)$ vectorials does not guarantee 
the presence of Higgs doublets in the low energy
massless spectrum. One has to assure the appropriate action 
of the $\alpha$ projections takes place, which is a time-consuming 
task from the point of view of model search.

There exists an elegant solution to the above problem that is 
related to a stringy doublet-triplet splitting mechanism. 
Moreover, it turns out that the 
relevant information, whether one of the triplets or the bi-doublet 
will survive, is encoded in each $SO(10)$ model; it does not depend 
on the GGSO projectors associated 
to the $SO(10)$ breaking $\alpha$ vector. In order to prove this we 
consider the action of the $2\alpha + x$ GGSO projection on the 
$SO(10)$ vectorial states of the
$V^{(A)}_{pqrs}$ sector, taking into account that 
$(2\alpha + x) \cap V^{(A)}_{pqrs}= \varnothing$: 
\begin{gather}
\nonumber \left[e^{i\pi (2\alpha+x)\cdot F_{V^A_{pqrs}}}- 
\delta_{V^A_{pqrs}}C\begin{pmatrix}
V^A_{pqrs}\\
2\alpha+x
\end{pmatrix}^*\right]\left\{\overline{\psi}^{1,2,3}_{1/2}\,,\,\overline{\psi}^{*1,2,3}_{1/2} \atop \overline{\psi}^{4,5}_{1/2}\,,\,\overline{\psi}^{4,5}_{1/2} \right\}\ket{\left(V^A_{pqrs}\right)_0}
= 0\\
\nonumber \Rightarrow \left[e^{i\pi \left[F(\overline{\psi}^{4}) + F(\overline{\psi}^{5})\right]} - C\begin{pmatrix}
V^A_{pqrs}\\
x
\end{pmatrix}\right]\left\{\overline{\psi}^{1,2,3}_{1/2}\,,\,\overline{\psi}^{*1,2,3}_{1/2} \atop \overline{\psi}^{4,5}_{1/2}\,,\,\overline{\psi}^{*4,5}_{1/2} \right\}\ket{\left(V^A_{pqrs}\right)_0}
= 0\\ 
\label{doublettriplet}
\Rightarrow C\begin{pmatrix}
V^A_{pqrs}\\
x
\end{pmatrix} = \begin{cases} 
-1 \ \leftrightarrow \ \text{the bidoublet survives}\\
+1 \ \leftrightarrow \ \text{the triplets survive}\\
\end{cases}
\end{gather}
In other words,  only $SO(10)$ vectorials originating from sectors with 
$C\left(V^A_{pqrs} \atop x\right) = -1$  could give rise to Higgs 
doublets. We call these states fertile vectorials. 
Their number, $N_{10}^f$, is given by
\begin{eqnarray}\label{bidoublet}
N_{10}^f = \frac{1}{2}\sum_{\substack{A=1,2,3 \\ p,q,r,s=0,1 }}\left[1 - 
C\left(V^A_{pqrs} \atop x\right)\right] R^{(A)}_{pqrs}
\end{eqnarray}
In general $N_{10}^f \le N_{10}$. As a minimal requirement a viable $SO(10)$
configuration should possess
\begin{align}
N^f_{10} \ge 1\,.
\label{VectorialNumber}
\end{align}
Nevertheless, the $2\alpha + x$ projection considered above is not 
completely equivalent to the $\alpha$ projection. The latter can in principle 
completely project out the bi-doublet even in the case where  
$C\binom{V^A_{pqrs}}{x} = -1$, so this fertility condition should be 
considered as necessary but not sufficient.

The advantage of this stringy doublet-triplet mechanism lies in the 
fact that it not only preserves the Higgs doublet pair but it also 
guarantees the absence of the associated triplet pair.
We should note that the above mentioned triplet representations are 
colour triplets, 
usually referred to as leptoquarks in the literature, which mediate 
proton decay via dimension five operators. Therefore, these states 
must be either sufficiently heavy so as to agree with the current 
proton lifetime of $\geq 10^{33}$ years \cite{proton} or must be 
projected out of the string spectrum by the GGSO projections.
The elegance of the string doublet-triplet mechanism has been previously 
noted, for example, in \cite{doublettriplet}, which works with NAHE-set 
based \cite{nahe} free fermionic models. In the NAHE models the 
doublet-triplet splitting occurs only in the untwisted sector, 
whereas here it can be applied to any twisted sector $SO(10)$ vectorial.

\subsection{Top Quark Mass Coupling}\label{topqconstraints}
In the class of models under consideration the top mass term 
stems from a superpotential coupling of the form
\begin{align}
Q_L\,Q_R\,h
\label{yukc}
\end{align}
where the left/right quarks and Higgs fields $Q_L, Q_R, h$ were 
defined in \eqref{sp_deco}, \eqref{vec_deco}. The conditions 
that assert the presence of this coupling at the tri-level 
superpotential were derived in \cite{tqmc}. The advantage of 
the formalism described in \cite{tqmc} is that it also fixes 
some of the degeneracy that appears in the free fermionic 
formulation (e.g. orbifold plane interchange). 
Without loss of generality we can choose that $Q_L$ arises from 
the sector $B^1_{0000} = S + b_1$, $Q_R$ comes from the sector 
$B^2_{0000} = S + b_2$, and $h$ comes from the sector 
$V^3_{0000} = S+ b_3 + x = S + b_1 + b_2$. In order for these 
states to survive the GGSO projections associated to 
$v_1,\dots,v_{12}$ vectors, the 
following conditions must be met
\begin{eqnarray}
& &C\left(b_1\atop e_1\right) = C\left(b_1\atop e_2\right) = 
C\left(b_1\atop z_1\right) = C\left(b_1\atop z_2\right) = +1\ ,\nonumber\\
\label{top_mass_coua}
& &C\left(b_2\atop e_3\right) = C\left(b_2\atop e_4\right) = 
C\left(b_2\atop z_1\right) = C\left(b_2\atop z_2\right) = +1\ ,\\
& &C\left(b_1\atop e_5\right) = C\left(b_2\atop e_5\right)\ , \ 
C\left(b_1\atop e_6\right) = C\left(b_2\atop e_6\right)~~
\nonumber\\
& & C\left(b_1\atop b_2\right) =  C\left(e_5\atop b_1\right) 
C\left(e_6\atop b_1\right)
\nonumber
\end{eqnarray}
In addition, the states that participate in \eqref{yukc} are subject 
to the $2\alpha +x$ GGSO projection. As explained in sections 3.1,3.2 
this projection is related to the $SU(2)_L\times{SU(2)}_R$
symmetry representations. Assuring the correct L/R transformation 
properties for $Q_L, Q_R$ translates to the 
additional constraints
\begin{align}
Q_L \ (S+b_1) \text{ survives } \iff & C\left(B^1_{0000}\atop x\right) = 
C\left(S + b_1 \atop x\right) = 1\nonumber \\
&\implies C\left(b_1 \atop x\right) = -1  \nonumber\\
Q_R \ (S+b_2) \text{ survives }  \iff & C\left(B^2_{0000}\atop x\right) = 
C\left(S + b_2 \atop x\right) = -1\nonumber \\
&\implies C\left(b_2 \atop x\right) = +1 \\
h (S+b_1+b_2) \text{ survives }\iff &C\left(V^3_{0000}\atop x\right) = 
C\left(S + b_1 + b_2 \atop x\right) = -1 \nonumber\\ 
&\implies  C\left(b_1 \atop x\right) C\left(b_2 \atop x\right) = -1 \nonumber
\end{align}
Only two of these constraints are independent
\begin{align}
\label{top_mass_coub}
C\left(b_1 \atop x\right) = -C\left(b_2 \atop x\right) = -1
\end{align}
and can be used to fix two additional GGSO coefficients, e.g. 
\begin{align}
c\left(\mathds{1}\atop b_1\right) &= -
c\left(e_3\atop b_1\right)
c\left(e_4\atop b_1\right)
c\left(e_5\atop b_1\right)
c\left(e_6\atop b_1\right)
\label{bccona}
\\
c\left(\mathds{1}\atop b_2\right) &= +
c\left(e_1\atop b_2\right)
c\left(e_2\atop b_2\right)
c\left(e_5\atop b_2\right)
c\left(e_6\atop b_2\right)
\label{bcconb}
\end{align}
Actually, the last two conditions are the doublet-triplet 
splitting constraints for the
vectorials stemming from the sector $S+b_3+x = S+b_1+b_2$.

Furthermore, the states that give rise to top quark mass coupling 
are subject to the GGSO projections related to the $SO(10)$ gauge 
symmetry breaking vector $v_{13} = \alpha$. Their survival is 
assured only in the case that two additional constraints are met
\begin{align}
c\left(b_1\atop \alpha\right) = c\left(b_2\atop \alpha\right) = -1
\label{alphacon}
\end{align}
From the technical point of view the above results have the 
advantage that they are explicit and consequently can be utilised to 
reduce the scanned parameter space.

\subsection{Fertile $SO(10)$ Cores}\label{fertileconditions}
Summarising the fertility conditions developed in sections 
\ref{osconstraints}-\ref{topqconstraints} can be done as follows:
\begin{enumerate}
\item Constraints related to the presence of a top quark mass coupling defined in Eqs. \eqref{top_mass_coua}, \eqref{top_mass_coub} and \eqref{bccona}, \eqref{bcconb}. 
These fix 13 entries in our $12\times 12$ matrix, leaving 55-13=42 independent phases, reducing the corresponding parameter space to $4.40\times 10^{12}$ $SO(10)$ 
string vacua. Moreover, Eq. \eqref{alphacon} fixes two of the additional twelve $\alpha$-vector related GGSO projections.
\item Constraints on $SO(10)$ spinorial states related to the presence of complete fermion families and $SU(2)_R\times{U(1)}_C$ symmetry breaking Higgs fields. For $n_g\ge 3$ these read
\begin{align}\label{fertility1}
 N_L-\overline{N}_L  = N_R-\overline{N}_R \geq 6 \ ,\ N_R > 6
\end{align}
\item Constraints related to the presence of the Standard Model breaking Higgs fields
\begin{align}\label{fertility2}
N^f_{10}\geq 1
\end{align}
\end{enumerate}
The above constraints do not guarantee the existence of phenomenologically promising LRS models, however, they result in a high likelihood
that such models will arise after employing the full GGSO projections on the massless string spectrum.
 We call $SO(10)$ models that comply with the above constraints ``fertile $SO(10)$ cores".
 
As explained, the first class of conditions can be expressed explicitly in terms of GGSO phases that define our parameter space. However, the second and third class of 
constraints cannot be explicitly solved in terms of GGSO phases. A scan of the related parameter space is required in order to extract  $SO(10)$ models that
satisfy these criteria. A comprehensive scan of the full parameter space, numbering  $4.40\times 10^{12}$ models, albeit straightforward,  requires considerable computer resources and computing time. 
It turns out that a random scan of the parameter space is quite efficient in capturing the salient phenomenological characteristics of these fertile cores. To this end we examine a sample of $10^9$ 
randomly selected configurations which corresponds to analysing one in one thousand models. A number of approximately 42000 fertile cores is collected through this procedure. 

As part of our methodology here we decided to incorporate an analysis of enhancements arising at the $SO(10)$ level and filtered out fertile cores which contained gauge group enhancements to the observable sector, whilst keeping those with no enhancements or enhancements affecting only the hidden gauge group factors. This procedure is described in the following section.

\subsection{Hidden Enhancements} \label{enhancements}
In the previous LRS classification \cite{LRSGlyn}, it was noted that 
approximately 29.1\% of LRS models contain additional gauge bosons 
but only the non-enhanced models were classified. 
In general, additional space-time vector 
bosons enhancing the gauge factors of (\ref{GG}) may arise from the following 26 sectors:
\begin{equation}\label{G}
\textbf{G} =
\begin{Bmatrix}
x & z_1 & z_2 & z_1 + z_2 \\
&&& \\
z_1 + 2\alpha & z_1 + z_2 + 2\alpha & 2\alpha + x & z_2 + 2\alpha + x \\
z_1 + 2\alpha + x & z_1 + z_2 + 2\alpha + x & & \\
&&& \\
\alpha & 3\alpha & z_1 + \alpha & z_1 + 3\alpha \\
z_2 + \alpha & z_2 + 3\alpha & z_1 + z_2 + \alpha & z_1 + z_2 + 3\alpha \\
\alpha + x & 3\alpha + x & z_1 + \alpha + x & z_1 + 3\alpha + x \\
z_2 + \alpha + x & z_2 + 3\alpha + x & z_1 + z_2 + \alpha + x & z_1 + z_2 +
3\alpha + x
\end{Bmatrix}
\end{equation}
where $x$ is defined in equation (\ref{x}).

However, in the current work, we are interested in exploring models with no enhancements or solely enhanced hidden sector gauge factors. Such hidden enhancements may arise 
from the sectors: $z_1,z_2$ and $z_1+z_2$. Such enhancements can be 
tested for at the $SO(10)$ level as they do not concern $\alpha$ 
GGSO phases and therefore fertile cores containing them can be 
found and included alongside non-enhanced cores in the analysis. In particular, after obtaining the approximately 42000 fertile cores from our scan of $10^9$ $SO(10)$ configurations, we then tested these cores for $SO(10)$ enhancements and filtered out those with observable enhancements and included cores within our sample with hidden enhancements. The hidden enhancement cases are presented in following tables. Note that in the tables we choose to use the arguments of the GGSOs:
\begin{equation}
e^{i\pi (v_i | v_j)}=C\binom{v_i}{ v_j }.  
\end{equation}
\begin{itemize}

\item $z_1+z_2=\{\bar{\phi}^{12345678}\}$ gives rise solely to spinorial 
hidden enhancementss.

\begin{center}
\begin{table}[ht]
\begin{tabular}{|c|c|}
\hline
\textbf{Enhancement Condition} & \textbf{Resulting Enhancement}\\
\hline
\makecell[l]{$(z_1+z_2|e_i)=(z_1+z_2|b_k )=0$\\
$(z_1+z_2|z_1)=1$} & \makecell{$SU(4)\times U(1)_4\times SU(2)\times 
U(1)_5\times U(1)_{7,8}\rightarrow$ \\ $SU(6)\times SO(4)\times U(1)$}\\
\hline
\makecell[l]{$(z_1+z_2|e_i)=(z_1+z_2|b_k )=0$\\
$(z_1+z_2|z_1)=0$} & \makecell{$SU(4)\times U(1)_4\times SU(2)\times 
U(1)_5\times U(1)_{7,8}\rightarrow$ \\ $SO(8)\times SU(2)\times SO(4) \times U(1)$}\\
\hline
\end{tabular}
\end{table}
\end{center}
\pagebreak
\item $z_1=\{\bar{\phi}^{1234}\}$ which gives rise to massless 
states of the form: $\psi^\mu_{\frac{1}{2}}\{\bar{y}^i,\bar{w}^i,
\bar{\psi}^{12345},\bar{\eta}^{123},\bar{\phi}^{5678}\}
\ket{\bar{\phi}^{1234}}$. In the following table however, only
the cases that result in enhancements to the hidden gauge group only are analysed, in particular the states: $\psi^\mu_{\frac{1}{2}}\{\bar{y}^i,\bar{w}^i,
\bar{\phi}^{5678}\}
\ket{z_1}$ are analysed.  

 \begin{center}
\begin{table}[ht]
\begin{tabular}{|c|c|}
\hline
\textbf{Enhancement Condition} & \textbf{Resulting Enhancement}\\
\hline
\makecell[l]{$(z_1|e_i)=(z_1|b_k)=(z_1|z_1)=0$ \\
$(z_1|z_2)=1$} & \makecell{$SU(4)\times U(1)_4\times 
SU(2)\times U(1)_5\times  U(1)_{7,8}\rightarrow $ \\ $SO(8)\times SO(4)\times SU(2)\times U(1)$}\\
\hline
\makecell[l]{$(z_1|e_i)=(z_1|b_k)=(z_1|z_2)=0$ \\
$(z_1|z_1)=1 $} & \makecell{$SU(4)\times U(1)_4\times SU(2)\times 
U(1)_5\times  U(1)_{7,8}\rightarrow$ \\ $SU(6)\times U(1)^3$}\\
\hline
\makecell[l]{$(z_1|e_j)=(z_1|z_1)=(z_1|z_2)=0$\\
$(z_1|\alpha)=0$ \\
$(z_1|e_i)=1$ \\
AND \\ 
~~\, $(z_1|b_1)=0, \ \ \ i=1,2$\\
or $(z_1|b_2)=0, \ \ \ i=3,4$\\
or $(z_1|b_1)=(z_1|b_2), \ \ \ i=5,6$\\} & 
\makecell{$SU(4)\times U(1)_4\times SU(2)\times U(1)_5\times  
U(1)_{7,8} \rightarrow$ \\ $SU(4)\times SO(3)\times SU(2)\times U(1)^3$}\\
\hline
\makecell[l]{$(z_1|e_j)=(z_1|z_1)=(z_1|z_2)=0$ \\
$(z_1|e_i)=(z_1|\alpha)=1$ \\
AND \\ 
~~\, $(z_1|b_1)=0, \ \ \ i=1,2$\\
or $(z_1|b_2)=0, \ \ \ i=3,4$\\
or $(z_1|b_1)=(z_1|b_2), \ \ \ i=5,6$\\} & 
\makecell{$SU(4)\times U(1)_4\times SU(2)\times U(1)_5\times  
U(1)_{7,8}\rightarrow$ \\ $SO(7)\times SU(2)\times U(1)^4$}\\
\hline
\end{tabular}
\end{table}
\end{center}
$i,j=1,...,6$, $i\neq j$, $k=1,2$.
\pagebreak
\item $z_2=\{\bar{\phi}^{5678}\}$ which gives rise to massless
states of the form: $\psi^\mu_{\frac{1}{2}}
\{\bar{y}^i,\bar{w}^i,\bar{\psi}^{12345},
\bar{\eta}^{123},\bar{\phi}^{1234}\}\ket{z_2}$. 
In the following table, the cases that result in enhancements to the 
hidden gauge group only are analysed, which are states of the form: $\psi^\mu_{\frac{1}{2}}\{\bar{y}^i,\bar{w}^i,
\bar{\phi}^{1234}\}
\ket{z_2}$.

\begin{table}[ht]
\begin{tabular}{|c|c|}
\hline
\textbf{Enhancement Condition} & \textbf{Resulting Enhancement}\\
\hline
\makecell[l]{$(z_2|e_i)=(z_2|b_k)=0$ \\
$(z_2|z_1)=1$} & \makecell{$SU(4)\times U(1)_4\times SU(2)\times U(1)_5\times  
U(1)_{7,8}\rightarrow$ \\ $SU(6) \times SO(4) \times U(1)$}\\
\hline
\makecell[l]{$(z_2|e_j)=(z_2|z_2)=(z_2|z_1)=0$ \\
$(z_2|e_i)=1$ \\
AND \\ 
~~\, $(z_2|b_1)=0, \ \ \ i=1,2$\\
or $(z_2|b_2)=0, \ \ \ i=3,4$\\
or $(z_2|b_1)=(z_2|b_2), \ \ \ i=5,6$\\} 
& \makecell{$SU(4)\times U(1)_4\times SU(2)\times U(1)_5\times  U(1)_{7,8}\rightarrow$ \\ $SU(4)\times U(1) \times SU(2)\times SO(3)^3$}\\
\hline
\end{tabular}
\end{table}
$i,j=1,...,6$, $i\neq j$, $k=1,2$.

\end{itemize}
Having filtered out observably enhanced cores, we were left with 19374 fertile cores free from observable $SO(10)$ enhancements. It's interesting to note that this means 53.9\% contain observable enhancements and so there's a higher correlation between these enhancements for fertile cores compared with a randomly generated $SO(10)$ cores. The origin of this correlation can be motivated by noting the appearance of constraints on the $(b_j|z_k)$, $j,k=1,2$, GGSO coefficients from the top quark mass coupling in equation (\ref{top_mass_coub}) coinciding with enhancement conditions.

The main characteristics of our remaining 19374 fertile cores,
including the number of generations ($n_g$), the number of spinorials/anti-spinorials that give rise to left and right states ($N_L/N_R$ and $\bar{N}^{}_L/\bar{N}^{}_R$) as well as the number of
vectorials that give rise to SM doublets, are presented in Table \ref{summarytable1}. 
\tiny
\begin{center}
\begin{table}[!htb]
\centering
\begin{tabular}{|c|c|c|c|c|c|c|}
\hline
$n_g$&$N^{}_L$&$\bar{N}^{}_L$&$N^{}_R$&$\bar{N}^{}_R$&$N^f_{10}$&frequency\\\hline
3&7 &1 &7&1& 2&       8060\\ \hline
3&6 &0 &8&2&6 &       3929\\ \hline
3&7 &1 &7&1& 4&    2796\\ \hline
3&6 &0 &8&2&2 &        1562\\ \hline
3&7 &1 &7&1&8 &        1247\\ \hline
3&6 &0 &8&2&10 &        497\\ \hline
4&8 &0 &10&2&8 &        450\\ \hline
4&10 &2 &10&2&2 &       232\\ \hline
4&9 &1 &9&1&4 &       212\\ \hline
4&8 &0 &12&4&8 &       124\\ \hline
4&10 &2 &10&2&4 &        86\\ \hline
4&8 &0 &12&4&16 &       60\\ \hline
4&8 & 2&8&2& 2&      38\\ \hline
4&10&2&10&2&8&        30\\ \hline
4&8&0&10&2&4&         27\\ \hline
4&8&2&8&2&8&          24\\ \hline

\end{tabular}
\caption{\label{summarytable1} \emph{Main characteristics of fertile $SO(10)$ cores with no observable enhancements for 19374 distinct models derived through a scan over $10^9$ randomly selected configurations (in a total of $4.4\times10^{12}$ possibilities)}}
\end{table}
\end{center}
\pagebreak
\normalsize
\section{Results and Analysis}\label{CRAL}

The use of fertile cores in our analysis means we are splitting the 
parameter space of LRS models into two components: $\Pi=\Pi_1\times \Pi_2 $. 
Where $\Pi_1$ is the space of $SO(10)$ models in which we select our 
fertile cores using our fertility conditions and the $\Pi_2$ subspace 
includes the GGSO phases related to the $SO(10)$ breaking vector 
$\alpha$.

As mentioned in section \ref{enhancements}, using a code 
written in Python we performed a scan over a random sample of $10^9$ vacua 
in the space $\Pi_1$ which consists of $4.4\times 10^{12}$ independent 
$SO(10)$ cores once the constraints from Section \ref{Fertility} are 
implemented. Cores satisfying the fertility 
conditions and containing no observable enhancements at the $SO(10)$ level are collected and in our sample 19374 fertile $SO(10)$ cores were found.

These 19374 cores are now to be explored in the LRS subspace $\Pi_2$ by iterating over 
its possible $\alpha$ GGSO coefficients. 
Considering equations (\ref{GSOdiagonals},\ref{alphacon}) there are in fact 9 independent $\alpha$ coefficients so each core results 
in $2^{9}=512$ LRS models which can be analysed and classified very quickly using our Python code.

Before analysing the results at the LRS level, we first present 
Figure 1 which displays how the number of vectorial bidoublets $N^f_{10}$ 
varies for 3 generation ($N_L-\bar{N}_L=6$, $N_R-\bar{N}_R=6$) cores from 
our sample of 19374 fertile cores. The results demonstrate that 3 
generation fertile cores come with $N^f_{10}=2,4,6,8$ or $10$ vectorial 
bidoublets and that $N^f_{10}=6$ is more common than $N^f_{10}=4$ in our sample. 
\begin{center}
\begin{figure}[!htb]
\centering
\includegraphics[width=0.85\linewidth]{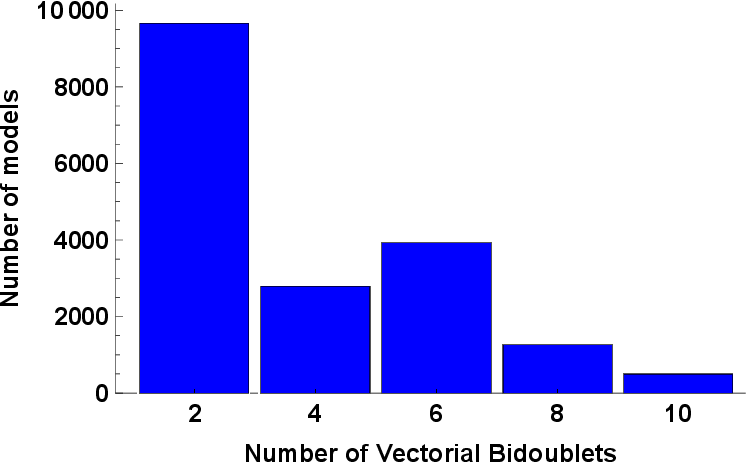}
\caption{Number of three generation fertile cores versus number of
twisted fertile vectorial representations from our set of 19374 fertile cores}
\end{figure}
\end{center}

The next step is the analysis of the LRS statistics resulting from our cores, which we obtain through a 
comprehensive scan of $\Pi_2$. The results are shown in Table \ref{Statstable}. These results ought to be 
compared and contrasted to the results of the
classification using the random classification method of \cite{LRSGlyn}, 
which are shown in a corresponding table on page 24 of ref. \cite{LRSGlyn} 
for a sample of $10^{11}$ LRS vacua.
\footnotesize
\begin{center}
\begin{table}[!htb]
\centering
\begin{tabular}{|c|l|r|c|r|}
\hline
&Constraints & \parbox[c]{2.5cm}{Total models in sample}& Probability \\
\hline
 & No Constraints & 9919488 & $1$  \\ \hline
(1)&{+ No Observable Enhancements} & 8894808 & $0.9$  \\  \hline
(2)& {+ No Chiral Exotics} & 1699104 & $0.17$  \\ \hline
(3)&{+ Complete Generations} & 1698818 & $0.17$  \\  \hline
(4)&{+ Three Generations} & 827333 & $8.3\times 10^{-2}$  \\  \hline
(5)&{+ SM Light Higgs}& 732728 & $7.4\times 10^{-2}$ \\
&{ \& Heavy Higgs}&&  \\\hline
(6)&{+ Top Quark Mass Coupling } & 732728 &  $7.4\times 10^{-2}$ \\ \hline
(7)&{+ Minimal Heavy Higgs} & 141568 &  $1.4\times 10^{-2}$ \\
&{ \& Minimal SM Light Higgs}&&  \\\hline
\end{tabular}
\caption{\label{Statstable} \emph{Statistics for the LRS models 
derived from fertile cores}}
\end{table}
\end{center}
\normalsize 

\hspace{0.67cm} The methodology just described is analogous to that used in the 
Standard-Like Model classification of \cite{SLM} except that here 
our comprehensive scan of $\Pi_2$ includes an analysis of the Enhanced,
Hidden and Exotic sectors too. As mentioned in section \ref{enhancements}, an important feature of our analysis was 
the inclusion of fertile cores which admitted an enhancement to the 
hidden sector gauge group. As can be seen in table \ref{Statstable} compared with Table 4 of \cite{LRSGlyn}, the probability of finding a model meeting all listed phenomenological criteria has increased from $4\times 10^{-11}$ to $1.4\times 10^{-2}$ due to our application of fertility conditions. This increase in probability by 9 orders of magnitude exhibits the power of the fertility methodology.

In order to explore the LRS statistics more closely, we can break down the results with respect to the important quantum numbers coming from the observable $\mathbf{16}$ representation and those coming from the vectorial $\mathbf{10}$ telling us the number of triplets and Higgs doublets at the LRS level. These statistics are presented in Table \ref{summarytable3} for three generation models in which the number of triplets, $n_3$, and anti-triplets, $n_{\bar{3}}$, are matched.


\section{Doublet-Triplet Splitting Discussion}\label{dtsd}
One notable result from our analysis is that there are no examples of good models that are also triplet-free. After noting this result, we ran another scan of the $SO(10)$ space $\Pi_1$ in which we added a further fertility constraint on top of those listed in Section \ref{fertileconditions} such that there are no triplets or anti-triplets at the $SO(10)$ level, which can be expressed through the equation:
\begin{eqnarray}\label{triplets}
N_{10}^t = \frac{1}{2}\sum_{\substack{A=1,2,3 \\ p,q,r,s=0,1 }}\left[1 - C\left(V^A_{pqrs} \atop x\right)\right] R^{(A)}_{pqrs}=0.
\end{eqnarray}
In our analysis we found that the only fertile cores with no triplets were found contained net chirality $N_L-\overline{N}_L=N_R-\overline{N}_R=8$ and so can only give rise to four generation models. Therefore, triplet--free models with four generations are likely to exist and performing a fishing algorithm for such four generation, triplet--free models allowed us to find models such as the following:

\begin{equation}
(v_i|v_j)= 
\begin{blockarray}{cccccccccccccc}
&\mathds{1}& S & e_1 & e_2 & e_3 & e_4 & e_5 & e_6 & b_1 & b_2 & z_1 & z_2 & \alpha \\
\begin{block}{c(ccccccccccccc)}
\mathds{1}&  1 &1 &1 &1 &0 &0 &1 &1 &1 &1 &1 &1 &-1/2   \\
S&  1 &1 &1 &1 &1 &1 &1 &1 &1 &1 &1 &1 &1  \\
e_1&  1 &1 &0 &0 &0 &0 &0 &0 &0 &1 &0 &0 &1  \\ 
e_2&  1 &1 &0 &0 &0 &0 &0 &0 &0 &1 &1 &0 &1   \\
e_3&  0 &1 &0 &0 &1 &0 &0 &1 &0 &0 &0 &0 &1   \\
e_4&  0 &1 &0 &0 &0 &1 &0 &0 &1 &0 &0 &0 &1 \\
e_5&  1 &1 &0 &0 &0 &0 &0 &0 &0 &0 &0 &0 &1   \\
e_6&  1 &1 &0 &0 &1 &0 &0 &0 &1 &1 &0 &0 &1   \\
b_1&  1 &0 &0 &0 &0 &1 &0 &1 &1 &1 &0 &0 &1   \\
b_2&  1 &0 &1 &1 &0 &0 &0 &1 &1 &1 &0 &0 &1   \\
z_1&  1 &1 &0 &1 &0 &0 &0 &0 &0 &0 &1 &1 &1   \\
z_2&  1 &1 &0 &0 &0 &0 &0 &0 &0 &0 &1 &1 &1   \\
\alpha&  1 &1 &1 &1 &1 &1 &1 &1 &0 &0 &0 &0 &1 \\
\end{block}
\end{blockarray}
\end{equation}
which is indeed triplet--free and four generation. This model also contains a top quark mass coupling, no chiral exotics, two heavy higgses, four standard model higgses and a hidden sector enhancement from the $z_2$ sector with oscillators $\{\bar{\phi}^{1234}\}$ enhancing the hidden group to $SU(6)\times SO(4)\times U(1)$. 

\tiny
\begin{center}
\begin{table}[!ht]
\centering
\begin{tabular}{|c|c|c|c|c|c|c|c|c|c|c|}
\hline
$Q_L$&$Q_R$&$L_L$&$L_R$&$\bar{Q}_L$&$\bar{Q}_R$&$\bar{L}_L$&$\bar{L}_R$&$N_h$&$n_3,n_{\bar{3}}$&\textbf{Frequency}\\ \hline
 3& 4& 3& 4& 0& 1& 0& 1& 3& 1 &    177152\\ \hline
 3& 4& 3& 4& 0& 1& 0& 1& 3& 3 &   167424\\ \hline
 3& 4& 3& 4& 0& 1& 0& 1& 1& 3 &   162304\\ \hline
 3& 4& 3& 4& 0& 1& 0& 1& 1& 5 &   91648\\ \hline
 3& 4& 3& 4& 0& 1& 0& 1& 1& 1 &   39424\\ \hline
 3& 4& 3& 4& 0& 1& 0& 1& 5& 1 &    38912\\ \hline
 3& 3& 4& 4& 0& 0& 1& 1& 0& 1 &    23488\\ \hline
 4& 4& 3& 3& 1& 1& 0& 0& 0& 1 &    23488\\ \hline
 3& 3& 4& 4& 0& 0& 1& 1& 2& 1 &    23488\\ \hline
 3& 4& 4& 3& 0& 1& 1& 0& 0& 1 &    23488\\ \hline
 4& 3& 3& 4& 1& 0& 0& 1& 0& 1 &    23488\\ \hline
 3& 4& 4& 3& 0& 1& 1& 0& 2& 1 &    23488\\ \hline
 4& 3& 3& 4& 1& 0& 0& 1& 2& 1 &    23488\\ \hline
 4& 4& 3& 3& 1& 1& 0& 0& 2& 1 &    23488\\ \hline
 4& 4& 3& 3& 1& 1& 0& 0& 2& 2 &    10624\\ \hline
 4& 3& 3& 4& 1& 0& 0& 1& 2& 2 &    10624\\ \hline
 3& 4& 4& 3& 0& 1& 1& 0& 2& 2 &    10624\\ \hline
 3& 3& 4& 4& 0& 0& 1& 1& 2& 2 &    10624\\ \hline
 3& 5& 3& 3& 0& 2& 0& 0& 1& 1 &    4864\\ \hline
 3& 3& 3& 5& 0& 0& 0& 2& 1& 1 &    4864\\ \hline
 3& 5& 3& 3& 0& 2& 0& 0& 3& 3 &    1792\\ \hline
 3& 3& 3& 5& 0& 0& 0& 2& 3& 3 &    1792\\ \hline
 4& 3& 3& 4& 1& 0& 0& 1& 0& 2 &     1328\\ \hline
 3& 3& 4& 4& 0& 0& 1& 1& 4& 2 &     1328\\ \hline
 3& 4& 4& 3& 0& 1& 1& 0& 0& 2 &     1328\\ \hline
 4& 4& 3& 3& 1& 1& 0& 0& 4& 2 &     1328\\ \hline
\end{tabular}
\caption{\label{summarytable3} \emph{LRS quantum number statistics for three generation models.}}
\end{table}
\end{center}
\normalsize
We note here that the doublet--triplet splitting mechanism, discussed 
in section \ref{ovsanddts}, 
involves the projection of the sector $(2\alpha+x)$, on the 
observable vectorial states arising in the vectorial sectors 
in eq. (\ref{ovsinovs}). The $(2\alpha+x)$ projection breaks the 
underlying $SO(10)$ GUT symmetry to the Pati-Salam subgroup. Hence, 
a doublet--triplet splitting mechanism, similar to the one that 
we discussed here, is operational in the Pati--Salam, as well as 
the Standard--like, heterotic--string models, that employ a 
Pati--Salam symmetry breaking basis vector in the construction. 

The absence of triplet-free three generation models in our sample 
may indicate that they are very rare, and hence are not generated in
our statistical sampling, or may result from a deeper reason in the 
structure of the LRS heterotic--string models. Three generation 
triplet-free models may also exist in the PS and SLM heterotic--string
models and can be searched for, by employing a similar analysis to that
of section \ref{ovsanddts}.

The fact that there are some four generation, triplet-free models though is somewhat reminiscent of the result for from the classification of FSU5 models in \cite{frs} that exophobic models exist only for even generation models. 

It is also worth reiterating here that all the models obtained by using
the free fermion classification method of refs \cite{fknr, fkr, acfkr,
frs, LRSGlyn} contain three pairs of vector--like triplet from the untwisted
Neveu--Schwarz sector, due to the fact that all these models utilise 
symmetric boundary conditions for the set of worldsheet fermions 
$\{y,\omega\vert {\bar y},{\bar\omega}\}^{1,\cdots,6}$. 

Projecting out the untwisted colour triplets and retaining 
the electroweak Higgs bidoublets, requires assignment of asymmetric 
boundary conditions for this set of worldsheet fermions, in the 
basis vector that breaks the $SO(10)$ symmetry to the Pati--Salam 
subgroup \cite{doublettriplet}. 
Implementation of the untwisted doublet-triplet splitting 
mechanism requires therefore extension of the classification
method to free fermion models with asymmetric boundary conditions.
Models which are free of both untwisted and twisted additional 
vector--like colour triplets may exist, but such models have
not been generated to date. 

Since the extra triplets appear in vector--like representation,
mass terms can be generated from cubic level and higher order
nonrenormalisable terms in the superpotential. In that case their
masses may be intermediate, rather than at the Planck scale. 
 
This situation is similar to that of the exotic fractionally 
charged states that are endemic in the heterotic--string models 
\cite{fcs}. A phenomenological requirement on such states
is that they appear in vector--like representations and are 
sufficiently massive or sufficiently rare to satisfy observational 
bounds. Models in which fractionally charged states appear only in the
massive string spectrum, but not among the massless physical states, 
were dubbed as exophobic string models. Exophobic 3 generation 
models were found
in the case of the PS heterotic--string models \cite{acfkr}, 
but not in the cases of the FSU5 \cite{frs}, the SLM \cite{SLM}, 
or the LRS \cite{LRSGlyn}, models. We may anticipate a similar 
situation with respect to the extra vector--like colour 
triplets that appear in these constructions. 

\section{Analysis of one Exemplary Model}\label{exmodel}
It is interesting to examine in detail one of our 
141568 exemplary models. 
As already mentioned, some of these models contain hidden enhancements but it's preferable to select a minimal model with no enhancement. We also choose a model with a minimal number of exotics states. It would have been preferable to have found a model with none of the vectorial triplets/antitriplets but, as already mentioned, no good models were found to derive from fertile cores with no triplet/anti-triplets. 

Using the notation convention:
\begin{equation}
C\binom{v_i}{ v_j } = e^{i\pi (v_i | v_j)}
\end{equation}
the model defined by the following 
GGSO projection coefficients is an example of such a minimal model:\\
\begin{equation}
(v_i|v_j)= 
\begin{blockarray}{cccccccccccccc}
&\mathds{1}& S & e_1 & e_2 & e_3 & e_4 & e_5 & e_6 & b_1 & b_2 & z_1 & z_2 & \alpha \\
\begin{block}{c(ccccccccccccc)}
\mathds{1}&1 & 1 &  0 &  0 &  1  & 0  & 1 & 1 & 1 & 0 & 1 & 1 & 1.5 \ \\  
S  & 1 & 1 & 1 & 1& 1 & 1 & 1 & 1 & 1 & 1 & 1 & 1 & 1  \  \\ 
e_1& 0 & 1 & 1 & 1 & 0 & 0 & 0 & 1 & 0 & 1 & 0 & 0 & 0 \ \\ 
e_2& 0 & 1 & 1 & 1 & 0 & 0 & 0 & 1 & 0 & 1 & 0 & 1 & 0 \ \\
e_3& 1 & 1 & 0 & 0 & 0 & 1 & 0 & 0 & 0 & 0 & 0 & 0 & 1 \ \\
e_4& 0 & 1 & 0 & 0 & 1 & 1 & 0 & 0 & 0 & 0 & 0 & 0 & 0 \ \\ 
e_5& 1 & 1 & 0 & 0 & 0 & 0 & 0 & 0 & 0 & 0 & 0 & 0 & 1 \ \\
e_6& 1 & 1 & 1 & 1 & 0 & 0 & 0 & 0 & 0 & 0 & 1 & 1 & 0 \   \\
b_1& 1 & 0 & 0 & 0 & 0 & 0 & 0 & 0 & 1 & 0 & 0 & 0 & 1 \  \\
b_2& 0 & 0 & 1 & 1 & 0 & 0 & 0 & 0 & 0 & 0 & 0 & 0 & 1 \   \\
z_1& 1 & 1 & 0 & 0 & 0 & 0 & 0 & 1 & 0 & 0 & 1 & 0 & 0 \  \\
z_2& 1 & 1 & 0 & 1 & 0 & 0 & 0 & 1 & 0 & 0 & 0 & 1 & 0 \  \\
\alpha & 1 & 1 & 0 & 0 & 1 & 0 & 1 & 0 & 0 & 0 & 1 & 1 & 1  \ \\
\end{block}
\end{blockarray}
\end{equation}
The observable matter sectors of this model produce three chiral generations, 
a top quark mass coupling, one SM Higgs and one heavy Higgs. There exists 
colour triplets from the vectorial \textbf{10} representation: one in the 
fundamental and one in the anti-fundamental, which pair up and obtain large 
mass. This model is free of chiral exotics and has 26 exotic states 
which is close to a minimum number of exotic states for LRS models since 
the classification done in \cite{LRSGlyn} found 22 exotic states as a lower bound on exotics. In particular, in the notation of Table 1 from \cite{LRSGlyn}, there are no chiral exotics since the spinorial exotic numbers are $n_{L_{L^s}}=1=n_{\bar{L}_{L^s}}$ and $n_{L_{R^s}}=1=n_{\bar{R}_{L^s}}$, whilst the vectorial exotic numbers are $n_{3v}=1=n_{\bar{3}v}$ and $n_{1v}=5=n_{\bar{1}v}$. Additionally, the Pati-Salam exotic numbers are $n_{L_{L^e}}=4$ and $n_{L_{R^e}}=10$. 

Another feature of this model is that it has an anomaly under the 
$U(1)_2$ and $U(1)_3$ gauge group factors since:
\begin{equation}
    \Tr U(1)_2=12 \ \ \ \text{and} \ \ \ \Tr U(1)_3 =12
\end{equation}
which results in an anomalous $U(1)$ combination of $$U(1)_A=U(1)_2+U(1)_3.$$
We note here that the existence and profile of the 
anomalous $U(1)$ in this model is in contrast to the case of the 
LRS NAHE--based models that were constructed in ref. \cite{lrs}.
In the NAHE models, the $U(1)_{1,2,3}$ were found to be anomaly free.
The reason is that the $\alpha$ projection selects opposite $U(1)_j$ 
charges for the left--handed and right--handed states from the 
sectors $b_j$. As a result, the sectors $b_j$ do not contribute to the
anomaly. The same holds in the LRS models that we analyse here using
our systematic classification method. The sectors $b_j$ therefore do not
contribute to the anomalous $U(1)$ also in the models that are 
generated by eq. (\ref{basis12}), and the contribution to the 
anomalous $U(1)$ arises from exotic states producing sectors.

\section{Conclusion}\label{conclusion}

The left--right symmetric models represent 
an appealing extension of the Standard Model \cite{lrsftmodel}
restoring the left--right symmetry in its spectrum, 
and attributing its violation to spontaneous symmetry breaking.
Furthermore, it mandates the existence of right--handed neutrinos and 
has a natural embedding in $SO(10)$. From the point of view of 
heterotic--string model building they also represent an interesting
case, as they do not follow from the more common $SO(12)\times E_8\times E_8$ 
route, but rather from the pattern $SO(16) \times E_7\times E_7$ \cite{lrs}. 
Resulting in models in which all $U(1)$ symmetries are anomaly free
\cite{lrs}, and in particular ${\rm Tr}U(1)_{1,2,3}=0$. In this paper, 
we presented a model with ${\rm Tr}U(1)_{1}={\rm Tr}U(1)_{2}\ne0$, 
in which case the contribution to the anomalies arises from 
exotic states producing sectors. 

In terms of the fermionic $Z_2\times Z_2$ classification program 
the LRS models present challenges that are similar to the SLM 
classification. 
In both cases there is a proliferation of 
exotic states producing sectors, lowering the frequency
of viable three generation models in the total space of 
models. For that purpose, one identifies fertile conditions 
at the $SO(10)$ level and selects cores that are amenable 
to producing three generation configurations. Around these
fertile cores a complete classification of the $SO(10)$ breaking 
phases is performed. In ref. \cite{LRSGlyn} a classification using the 
random generation method was performed producing a small number of 
three generation models. Adopting the two stage classification
method in this paper, the number of viable models is increased 
by four orders of magnitude. Furthermore, we showed that the
fertility conditions are associated with a novel doublet--triplet 
splitting mechanism that operates in the twisted sectors of the
LRS vacua. While a doublet--triplet splitting was demonstrated in the past
for untwisted states \cite{doublettriplet}, the doublet--triplet 
splitting mechanism identified herein operates in the twisted 
sectors, and may be employed in SLM and PS heterotic--string models 
as well. The stage is now ripe for adopting
novel computational methods in the classification program 
\cite{fhpr} to identify patterns in the GGSO coefficient space
that are conducive for producing viable phenomenological 
characteristics. 

\section*{Acknowledgments}

AEF would like to thank the Galileo Galilei Institute, CERN, the Simons 
Center, the Weizmann Institute, and Oxford University for hospitality,
where part of this work was conducted. The work of BP is supported in 
part by STFC grant ST/N504130/1.


\bibliographystyle{unsrt}

\end{document}